\documentclass[aps,prl,reprint,superscriptaddress]{revtex4-2}
\usepackage{url}
\usepackage{amssymb}
\usepackage{mathtools}
\usepackage{graphicx}
\usepackage{xcolor}

\date{\today}

\begin{document}
\title{Connectedness percolation of fractal liquids}
	\author{René de Bruijn}
	\email{r.a.j.d.bruijn@tue.nl}
	\affiliation{Department of Applied Physics, Eindhoven University of Technology, P.O. Box 513, 5600 MB Eindhoven, Netherlands}
	\author{Paul van der Schoot}
	\affiliation{Department of Applied Physics, Eindhoven University of Technology, P.O. Box 513, 5600 MB Eindhoven, Netherlands}
\begin{abstract}
We apply connectedness percolation theory to fractal liquids of hard particles, and make use of a Percus-Yevick liquid state theory combined with a geometric connectivity criterion. We find that in fractal dimensions the percolation threshold interpolates continuously between integer-dimensional values, and that it decreases monotonically with increasing (fractal) dimension. The influence of hard-core interactions is only significant for dimensions below three. Finally, our theory incorrectly suggests that a percolation threshold is absent below about two dimensions, which we attribute to the breakdown of the connectedness Percus-Yevick closure.
\end{abstract}
\maketitle

Recently, Heinen \textit{et al.} introduced \textit{fractal liquids} in which both the particles and the embedding space are treated as objects of the same fractal dimension \cite{Heinen2015}. Such liquids are therefore fractal at all length scales. This contrasts with the more familiar case of fluids confined in porous media, which are often thought to represent a fractal geometry. In liquid-state theory, the \textit{local} structure of the confining medium is in that case usually modeled as a sphere, cylinder or slit \cite{Li2016, Ji2009}, and any connection to the fractal background lost. Exceptions are so-called quenched-annealed liquids of which the constituent model particles share the volume with confining obstacles the distribution of which is fixed in space \cite{Madden1988,given1992,Schmidt2005}.

Confinement is known to have a significant impact on phase transitions, \textit{e.g.}, by shifting the critical point, changing the order of the phase transition, or even causing a phase transition to be absent altogether \cite{Chan1988,Dadmun1993,Brown1998,Iannacchione1993,Kityk2008,Erdal2012}.
This is mirrored, on the one hand, in theoretical studies of phase transitions in cylinders \cite{Maddox1997,Gelb1999Review,Alba2006Review} and slits \cite{delasHeras2005,Erdal2012,Gelb1999Review,Alba2006Review}, and, on the other hand, by those that \textit{effectively} describe the actual structure of a porous medium.
In the latter, the fractal geometry is either inscribed explicitly in a lattice \cite{Gefen1980,*Gefen1983,*Gefen1984a,*Gefen1984,Windus2009} or treated implicitly by a random disorder field in continuum field theories \cite{Popa-Nita2001,Radzihovsky1999,Feldman2000,Hvozd2018}.  

In the theory of fractal liquids, however, the porosity of the confining medium is described by a single (fractal) dimension, so without any reference to a Euclidean embedding space, and specific interactions with the confining walls are ignored \cite{Heinen2015}.
The predictions of Heinen \textit{et al.} for the microscopic fluid structure, obtained using a generalized Percus-Yevick approach, agree very well with results from their Monte Carlo simulations \cite{Heinen2015}. Actual realizations of this model may perhaps be found in binary microphase-separated liquids in porous media if the characteristic size of the (macroscopic) liquid droplets is very much larger than the porosity length scale. 

As far as we are aware, phase transitions in fractal liquids have not yet been investigated. Hence, in this Letter, we focus attention on the geometric percolation transition in fractal liquids, which belongs to a particular class of (second order) phase transition \cite{Torquato2002}. 
Of particular interest is the influence of the fractal dimension $D$ on the percolation threshold, defined as the filler fraction at which a material-spanning cluster emerges, and the critical exponent $\gamma$, associated with the mean cluster size.
To calculate these quantities, we make use of the Percus-Yevick integral equation theory for fractal liquids of Heinen and collaborators \cite{Heinen2015}, and apply it to geometrical percolation where connectivity is defined by a distance criterion. 
In our so-called cherry-pit model, the particles have an impenetrable core of diameter $\sigma$ and direct connections are identified by this distance criterion $\lambda$. 
In principle, both the percolation threshold and the critical exponent $\gamma$ depend on the ratio $\sigma/\lambda$. 
As far as we are aware, cPY theory within the cherry-pit model has only been analyzed in $D = 3$ \cite{DeSimone1986}, hence our analysis extends to both integer and non-integer dimensions between one and six.

According to our findings, the geometric percolation threshold of fractal liquids of hard particles interpolates in a continuous manner between those of integer-dimensional fluids of isometric particles, and decreases monotonically with increasing fractal dimension. The critical exponent $\gamma$ also decreases with increasing dimensionality and approaches the mean-field value of unity already in five dimensions. This is below the accepted upper critical dimension of six \cite{Torquato2002}. Surprisingly, our calculations indicate that connectedness Percus-Yevick theory breaks down approaching two dimensions from above: the critical exponent $\gamma$ diverges for $D \downarrow 2$ and in that case we fail to find a system-spanning cluster at finite densities. 
Interestingly, we find that the value of $\sigma/\lambda$ either weakly impacts upon our findings or not at all.

\begin{figure}[bt]
	\centering
	\includegraphics[width=8.6cm]{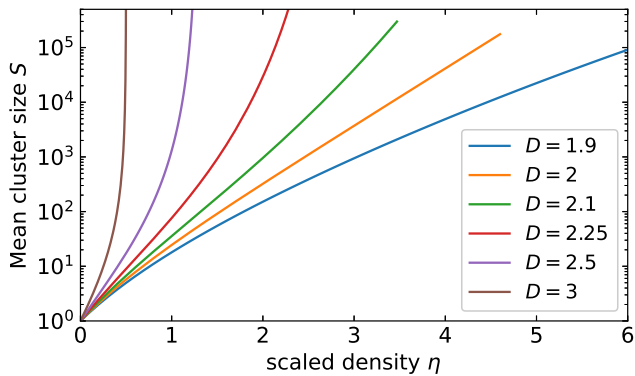}
	\caption{The mean cluster size $S(D)$ obtained within cPY theory as function of the scaled density $\eta$, defined as the number density scaled with the volume of a $D$-dimensional sphere of diameter $\lambda$.
}
	\label{fig:Svsphi}
\end{figure}

Postponing a discussion of our formalism we first highlight in more detail our findings on ideal, non-interacting fractal particles for which $\sigma/\lambda=0$. Fig.~\ref{fig:Svsphi} shows how according to our calculations the mean cluster size $S(D)$ of such particles depends on the scaled density $\eta \equiv 2 \pi^{D/2} (\lambda/2)^D \rho/D\Gamma(D/2)$, for selected dimensionalities $D$ between $1.9$ and $3.0$. Here, $\rho$ is the number density of the ideal particles of ``diameter'' $\lambda$, and $2 \pi^{D/2} (\lambda/2)^D/D\Gamma(D/2)$ the volume of a $D$-dimensional sphere with diameter $\lambda$. 

As the percolation threshold is the scaled density for which the mean cluster size $S$ diverges, we deduce from Fig.~\ref{fig:Svsphi} that within cPY theory this seems not to occur for $D \leq 2$. For $D = 2$, the mean cluster size grows exponentially up to the largest density of $4.6$ that in our calculations produce a convergent cluster size. This, incorrectly, suggests that for $D=2$ it formally diverges at an infinite density. In this context, it is useful to note that this should certainly happen for $D=1$. Indeed, the exact result for the cluster size in \textit{one} dimension reads $S(1) = 2\exp \eta-1$. If we compare this with the prediction of cPY theory, $S_\mathrm{PY}(1) = (1 + \eta)^2$, then it transpires that both remain finite at finite density but differ considerably in functional form \cite{Drory1997}. This calls into question the validity of cPY theory for $D \leq 2$.

\begin{figure}[bt]
	\centering
	\includegraphics[width=8.6cm]{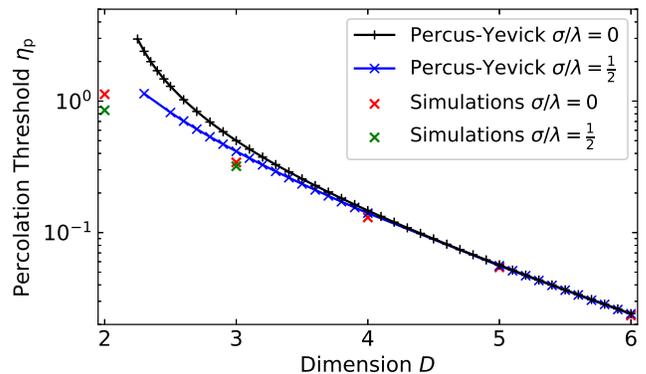}
	\caption{The percolation threshold $\eta_\mathrm{p}$ as function of the spatial dimension $D$ for cherry-pit particles with $\sigma/\lambda = 0$ (black, plusses) and $\sigma/\lambda = 1/2$ (blue, crosses). The percolation threshold is expressed as the number density scaled with the volume of a $D$-dimensional sphere of diameter $\lambda$. Results obtained from Monte Carlo simulations for $\sigma/\lambda = 0$ (red) are taken from Ref.~\cite{TorquatoJiao2012}, and for $\sigma/\lambda = 1/2$ from Ref.~\cite{Lee1990,Miller2009} (green).}
	\label{fig:ptvsD}
\end{figure}

The obvious question that now arises is how well cPY fares for $D > 2$. Fig.~\ref{fig:Svsphi} suggests that for $D > 2$ the mean cluster size diverges at a finite density. Indeed, the analytical solution of cPY theory for $\sigma/\lambda=0$ in the \textit{integer} dimension $D = 3$ gives a percolation threshold of $\eta=\eta_\mathrm{p} = 1/2$. For $D = 5$, we find $\eta = 3/2 - 5/6 \sqrt{3}$. The former overestimates Monte Carlo simulation results \cite{Lorenz2001a} by almost 50\%, whilst the latter overestimates Monte Carlo results by about 4\% \cite{TorquatoJiao2012}. In Fig.~\ref{fig:ptvsD} we show our numerically obtained percolation threshold for the cases $\sigma/\lambda =0$ and $0.5$ as function of the (fractal) dimension $D$, and compare these with simulation results for integer dimensions $D = 2 - 6$. It shows that the presence of a hard core does not appreciably affect the percolation threshold. See also the Supplemental Material \cite{SI}.

For $D < 3$, the percolation threshold increases sharply with decreasing dimension, and appears to diverge upon approach of $D \downarrow 2$, although we have not been able to extract the percolation threshold for $D < 2.25$. This supports our previous assessment based on Fig.~\ref{fig:Svsphi}. For integer $D \geq  4$, theory and simulations agree almost quantitatively, with the percolation threshold decreasing with increasing dimension. A decreasing $\eta_\mathrm{p}$ with increasing $D$ is to be expected if we take the percolation threshold to be inversely proportional to the volume available for two particles to remain connected \cite{Balberg1984,Torquato2012,TorquatoJiao2012}. To leading order this gives $\eta_\mathrm{p} \propto 2^{-D}$, which becomes exact in infinite dimensions \cite{Torquato2012,Grimaldi2015}.

Having presented our main findings for the percolation threshold of fractal particles, we now describe our formalism and after that discuss in more detail the subtle influence of $\sigma/\lambda$, and that of dimensionality, on the critical exponent. Before going into the details of our calculations, it seems sensible to first introduce two concepts that are relevant in the context of the fractal nature of our particles, and the space they live in. 

The first point we need to address, is that the relevant distance measure is \textit{not} the Euclidean but the so-called \textit{chemical} distance, where the distance between two points is measured along the fractal embedding space \cite{Heinen2015}. In lattice terminology, this translates to the shortest connected path between two sites \cite{Heinen2015}. Further, the relevant (fractal) dimension in the model identified by Heinen and co-workers is the \textit{spreading dimension} $d_l$. It is related to the chemical distance by the scaling of the number of sites (or ``mass") $\mathcal{N}$ that are within the chemical distance $l_{\mathrm{chem}}$ from any site via $\mathcal{N} \sim l_{\mathrm{chem}}^{d_l}$ \cite{Heinen2015}. In integer dimensions, where the chemical distance coincides with the Euclidean distance, the spreading dimension coincides with the spatial dimension. 

With these definitions, we can now generalize our cherry-pit particle model to fractal dimensions. In lattice terminology, we define the fractal dimensional equivalent of a hard core particle with ``diameter'' $\sigma$, as all nodes that lie within a chemical distance of $\sigma/2$ removed from the center node. Moreover, the fractal particles have a connectivity shell of diameter $\lambda$ around this hard core. If the chemical distance between the centers of two particles is less than $\lambda$, yet larger than $\sigma$, we define the particles to be connected. Due to the hard core repulsion, the centers of two particles cannot be within a chemical distance of $\sigma$. 

Our theoretical description of geometric percolation is based on connectedness Ornstein-Zernike (cOZ) theory \cite{Coniglio1977}. Within this formalism, the cluster size is given by $S = 1 + \rho \lim\limits_{q \to 0} \widehat{P}(q)$, where $\rho$ is the number density and $\widehat{P}(q)$ is the Fourier Transform of the so-called pair connectedness function $P(r)$. The pair connectedness function describes the probability that two particles, separated by a center-to-center distance $r = |\mathbf{r}|$, are connected.
It is connected to the function $C^+(r)$ known as the direct connectedness function, via the cOZ equation
$ P(r) = C^+(r) + \rho\int \mathrm{d}^{D}\mathbf{r}'P(r') C^+(|\mathbf{r}-\mathbf{r}'|)$, with
$D$ again the spreading dimension, $C^+(r)$ encoding the specific subset of connections between pairs of particle that remain connected upon removal of any other particle connected to these two  \cite{Coniglio1977}. 

Obviously, since $C^+(r)$ is unknown \textit{a priori}, the cOZ equation needs to be supplemented by a closure relation. We employ the connectedness Percus-Yevick or cPY closure, defined by the conditions $P(r \leq \lambda) = g(r)$, and $C^+(r > \lambda) = 0$ \cite{DeSimone1986}. The latter imposes the presumed short-distance nature of the direct connectedness function. That the former is sensible follows from the fact that the radial distribution function $g(r)$ describes the probability to find a particle at $r$ around such a test particle placed at the origin. 
Our main motivation for using the cPY closure is that it allows us to obtain analytical results for ideal particles in odd dimensions. 

The radial distribution function itself can be obtained from the liquid-state Ornstein-Zernike (OZ) equation $g(r) = 1 + c(r) + \rho\int \mathrm{d}^{D}\textbf{r}'\left[g(|\textbf{r}'|)-1\right] c(|\textbf{r} - \textbf{r}'|)$, which also needs to be closed. As we use the Percus-Yevick closure for the cOZ equation we invoke the same closure here, implying that for hard particles we insist on the no-overlap condition $g(r \leq \sigma) = 0$ and set $c(r > \sigma) = 0$  \cite{Hansen2013}. 
For ideal particles $g(r) = 1$ for all $r \geq 0$, and only the cOZ equation needs to be solved, which we do numerically, simplifying our calculations considerably. 
For cherry-pit particles with $\sigma/\lambda > 0$, we numerically solve the OZ and cOZ equations consecutively, and rely on the same method used by Heinen and co-workers, that is, by exploiting a generalized Hankel transform that can be dimensionally continued (see Supplemental Material \cite{SI}) \cite{Heinen2015}.

Finally, we pinpoint the particle density at the percolation threshold $\rho_\mathrm{p}$, or in dimensionless form $\eta_\mathrm{p}$, by the condition $S \to \infty$. This we also do numerically, making use of the scaling relation for the mean cluster size $S \propto |\eta-\eta_\mathrm{p}|^{-\gamma}$
presumed to be valid for $\eta \to \eta_\mathrm{p}$. Here,  $\gamma$ is the appropriate critical exponent. The quantities $\eta_\mathrm{p}$ and $\gamma$ we asymptotically fit in the critical region of the mean cluster size $S$. 
We have tested this procedure, and compare it against the exact analytical results for ideal particles in $D = 3$ and $5$, and find the error in the percolation threshold $\eta_\mathrm{p}$ to be negligible (less than $10^{-2}$\%). The error in the critical exponent $\gamma$ is somewhat larger, up to four percent from the analytically obtained values. We refer to the Supplemental Material for a detailed discussion \cite{SI}. 

\begin{figure}[bt]
	\centering
	\includegraphics[width=8.6cm]{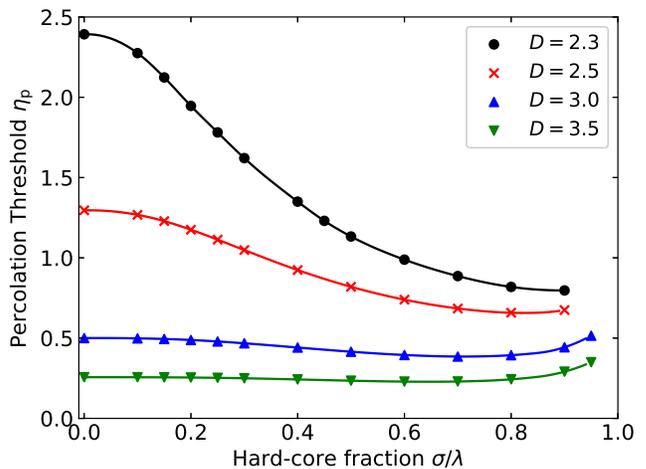}
	\caption{The percolation threshold $\eta_\mathrm{p}$, expressed as the number density scaled with the volume of a $D$-dimensional sphere of diameter $\lambda$, as function of the hard core fraction $\sigma/\lambda$ for spatial dimensions $D = 2.3$ (dots), $2.5$ (crosses), $3.0$ (triangle up) and $3.5$ (triangle down). Here, $\sigma$ is the hard core diameter, and $\lambda$ the diameter of the connectivity shell. The lines are a spline fit through the data as a guide for the eye.}
	\label{fig:ptvsf}
\end{figure}

We present results of our calculations for cherry-pit particles in Fig.~\ref{fig:ptvsf}, showing the percolation threshold as function of the hard-core fraction $\sigma/\lambda$ for selected dimensions. We restrict ourselves to those results for which we can pinpoint the percolation threshold accurately, that is, for $D > 2.25$. We notice that, starting at $\sigma/\lambda = 0$, the percolation threshold decreases with increasing $\sigma/\lambda$ albeit that the effect is larger the smaller the dimensionality of space. However, for $D\geq 2.5$, we find that the percolation threshold increases again, \textit{i.e.}, there is a well-defined minimum for some value of  $\sigma/\lambda > 0$ that depends on the value of $D$.

For $D=2$ and $3$, this non-monotonic behavior can be explained in terms of two counteracting many-body effects \cite{Bug1985}. 
The first is connected with that fewer particles are, on average, required to span a certain distance in the presence of a hard core, and moreover these configurations are more probable due to local crowding of particles around that hard core. This effect \textit{decreases} the percolation threshold. The second effect is caused by the connectivity shell becoming smaller with increasing value of $\sigma/\lambda$. The concomitant decrease in contact volume \textit{increases} the percolation threshold. The former effect predominates more strongly in lower dimensional spaces, because the available ``volume'' per particle decreases with decreasing dimensionality. 

Of the findings presented in Fig.~\ref{fig:ptvsf}, only those for $D=3$ allow for comparison with Monte Carlo simulations reported on in the literature \cite{Miller2009}. As is well-known, cPY predictions deviate by approximately $46\%$ for $\sigma/\lambda = 0$, but the difference decreases with increasing $\sigma/\lambda$ down to $14\%$ for $\sigma/\lambda = 0.95$. Incidentally, for $\sigma/\lambda > 0.95$ percolation is preempted then by a transition to a crystal phase \cite{DeSimone1986,Miller2009}. If we stay below the crystal transition, we expect cPY to be most accurate for small connectivity ranges for all $D>1$, not just $D = 3$. The reason is that with increasing $\sigma/\lambda$, the cluster structure becomes increasingly more tree-like \cite{Grimaldi2017b}. Nevertheless, the observation from Fig.~\ref{fig:ptvsD} that the theory becomes less accurate for $D < 3$ generalizes for all $0\leq \sigma/\lambda \leq 1$.

\begin{figure}[bt]
	\centering
	\includegraphics[width=8.6cm]{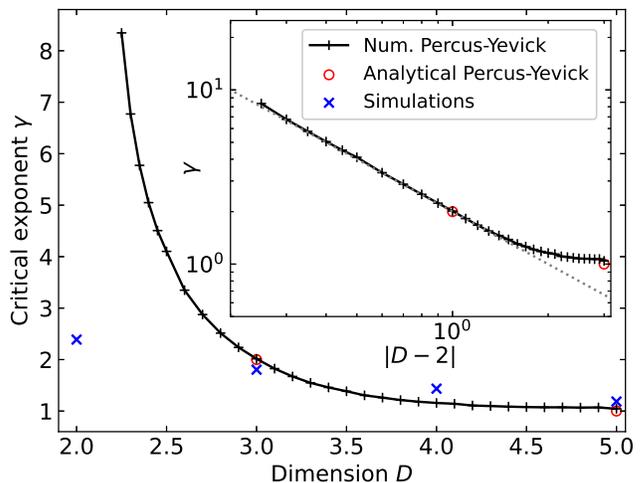}
	\caption{Main: The critical exponent $\gamma$ within cPY theory both numerically (black, plusses) and theoretically (red, circles), and from simulations (blue, crosses) \cite{DeSimone1986,Adler1990}. Inset: The critical exponent $\gamma$ as function of the shifted dimension $|D-2|$, where including the scaling $\gamma \sim 2/(D-2)$ (grey, dotted). 
	}
	\label{fig:gammavsdim}
\end{figure}

Taking cPY at face value for all $D$ and $
\sigma/\lambda$, then both Fig.~\ref{fig:ptvsD} and Fig.~\ref{fig:ptvsf} lead us to the conclusion that the percolation threshold must rise substantially upon approaching two dimensions from above. 
Associated with this apparent divergence in the percolation threshold, we find a divergence of the critical exponent $\gamma$. Our most accurate estimate for $\gamma$ we obtain for the case $\sigma/\lambda = 0$, and is presented in Fig.~\ref{fig:gammavsdim}. We do not expect that a non-zero $\sigma/\lambda$ changes this as the cherry-pit and ideal models should be in the same universality class \cite{Chiew1983,DeSimone1986}. Representative findings for $\sigma / \lambda > 0$, presented in the Supplemental Material, support this \cite{SI}. 

As is evident from Fig.~\ref{fig:gammavsdim}, the critical exponent interpolates continuously between the known cPY exponent in three dimensions $\gamma = 2$ and the exponent $\gamma = 1$ obtained by us for $D=5$ (See Supplemental Material \cite{SI}). It shows the same trend as the results from Monte Carlo simulations, also indicated, where $\gamma$ increases with decreasing value of $D$. We note that the critical exponent we find for $D =5$ is the mean-field value, yet the generally accepted upper critical dimension  for both lattice and continuum percolation is $D = 6$ \cite{Torquato2002}. 

The sharp rise of the critical exponent when the dimensionality of space drops below three contrasts with the simulation results.
In the inset of Fig.~\ref{fig:gammavsdim} we suggest that $\gamma$ scales as $\gamma = 2/(D-2)$ for $2 < D < 3$, which indeed points at $\gamma$ diverging for $D \to 2$. Incidentally, a similar divergence is known to occur in the  spherical model of ferromagnetism \cite{Baxter1982}. This strengthens our conclusion that cPY theory breaks down near $D=2$.

It is not clear exactly why cPY theory fails near two dimensions. Of course, we cannot exclude the possibility that it is not cPY theory itself that lies at the root of the problem but some numerical issue. Still, it should not come as a complete surprise, because 
percolation is essentially a high-density phenomenon as Fig.~\ref{fig:ptvsD} also shows. For \textit{penetrable} particles, the actual fraction of the volume covered by particles at the percolation threshold is $\phi_\mathrm{p} = 1 - \exp(-\eta_\mathrm{p}) \approx 0.67$ in two dimensions compared to $\phi_\mathrm{p} \approx 0.28$ in three dimensions and to $\phi_\mathrm{p} \approx 0.12$ in four \cite{Torquato2002}. It follows that the long-ranged loop connections that cPY theory neglects must become increasingly important when lowering the dimensionality of space \cite{Coupette2021}. There is no reason to suspect this not also to be true for hard particles \cite{Lee1990}. As is becoming increasingly clear, closures that are accurate in the context of thermodynamic liquid-state theory are not necessarily accurate in the context of percolation, in particular in low dimensional systems, \cite{Coupette2020}, and that they have to be adapted for that purpose \cite{Coupette2021}.

In conclusion, we have investigated the geometrical percolation transition in fractal liquids within a cherry-pit model, and applied for that the Percus-Yevick approximation. We find that the continuum percolation threshold in non-integer dimensions interpolates continuously between the integer-dimensional values, and decreases with increasing dimension. The same conclusion holds for the critical exponent $\gamma$, which within Percus-Yevick theory attains its mean-field value in five dimensions, below the generally accepted upper critical dimension of six. Interestingly, hard-core interactions affect the percolation threshold only marginally, in particular in higher-dimensional spaces.
Below three dimensions, the percolation threshold $\eta_\mathrm{p}$ as well as the critical exponent  $\gamma$ diverge as $D \to 2$. This contrasts with the known finite percolation threshold and critical exponent for $D=2$, and signifies the breakdown of connectedness Percus-Yevick theory below three dimensions. 

\begin{acknowledgments}
R.d.B. and P.v.d.S. acknowledge funding by the Institute for Complex Molecular Systems at Eindhoven University of Technology.
\end{acknowledgments}

\bibliography{bibliography}

\begin{thebibliography}{45}%
\makeatletter
\providecommand \@ifxundefined [1]{%
 \@ifx{#1\undefined}
}%
\providecommand \@ifnum [1]{%
 \ifnum #1\expandafter \@firstoftwo
 \else \expandafter \@secondoftwo
 \fi
}%
\providecommand \@ifx [1]{%
 \ifx #1\expandafter \@firstoftwo
 \else \expandafter \@secondoftwo
 \fi
}%
\providecommand \natexlab [1]{#1}%
\providecommand \enquote  [1]{``#1''}%
\providecommand \bibnamefont  [1]{#1}%
\providecommand \bibfnamefont [1]{#1}%
\providecommand \citenamefont [1]{#1}%
\providecommand \href@noop [0]{\@secondoftwo}%
\providecommand \href [0]{\begingroup \@sanitize@url \@href}%
\providecommand \@href[1]{\@@startlink{#1}\@@href}%
\providecommand \@@href[1]{\endgroup#1\@@endlink}%
\providecommand \@sanitize@url [0]{\catcode `\\12\catcode `\$12\catcode
  `\&12\catcode `\#12\catcode `\^12\catcode `\_12\catcode `\%12\relax}%
\providecommand \@@startlink[1]{}%
\providecommand \@@endlink[0]{}%
\providecommand \url  [0]{\begingroup\@sanitize@url \@url }%
\providecommand \@url [1]{\endgroup\@href {#1}{\urlprefix }}%
\providecommand \urlprefix  [0]{URL }%
\providecommand \Eprint [0]{\href }%
\providecommand \doibase [0]{https://doi.org/}%
\providecommand \selectlanguage [0]{\@gobble}%
\providecommand \bibinfo  [0]{\@secondoftwo}%
\providecommand \bibfield  [0]{\@secondoftwo}%
\providecommand \translation [1]{[#1]}%
\providecommand \BibitemOpen [0]{}%
\providecommand \bibitemStop [0]{}%
\providecommand \bibitemNoStop [0]{.\EOS\space}%
\providecommand \EOS [0]{\spacefactor3000\relax}%
\providecommand \BibitemShut  [1]{\csname bibitem#1\endcsname}%
\let\auto@bib@innerbib\@empty
\bibitem [{\citenamefont {Heinen}\ \emph {et~al.}(2015)\citenamefont {Heinen},
  \citenamefont {Schnyder}, \citenamefont {Brady},\ and\ \citenamefont
  {L{\"{o}}wen}}]{Heinen2015}%
  \BibitemOpen
  \bibfield  {author} {\bibinfo {author} {\bibfnamefont {M.}~\bibnamefont
  {Heinen}}, \bibinfo {author} {\bibfnamefont {S.~K.}\ \bibnamefont
  {Schnyder}}, \bibinfo {author} {\bibfnamefont {J.~F.}\ \bibnamefont
  {Brady}},\ and\ \bibinfo {author} {\bibfnamefont {H.}~\bibnamefont
  {L{\"{o}}wen}},\ }\href {https://doi.org/10.1103/PhysRevLett.115.097801}
  {\bibfield  {journal} {\bibinfo  {journal} {Physical Review Letters}\
  }\textbf {\bibinfo {volume} {115}},\ \bibinfo {pages} {097801} (\bibinfo
  {year} {2015})}\BibitemShut {NoStop}%
\bibitem [{\citenamefont {Li}\ \emph {et~al.}(2016)\citenamefont {Li},
  \citenamefont {Suen}, \citenamefont {Prince}, \citenamefont {Larin},
  \citenamefont {Klinkova}, \citenamefont {Th{\'e}rien-Aubin}, \citenamefont
  {Zhu}, \citenamefont {Yang}, \citenamefont {Helmy}, \citenamefont
  {Lavrentovich} \emph {et~al.}}]{Li2016}%
  \BibitemOpen
  \bibfield  {author} {\bibinfo {author} {\bibfnamefont {Y.}~\bibnamefont
  {Li}}, \bibinfo {author} {\bibfnamefont {J.~J.-Y.}\ \bibnamefont {Suen}},
  \bibinfo {author} {\bibfnamefont {E.}~\bibnamefont {Prince}}, \bibinfo
  {author} {\bibfnamefont {E.~M.}\ \bibnamefont {Larin}}, \bibinfo {author}
  {\bibfnamefont {A.}~\bibnamefont {Klinkova}}, \bibinfo {author}
  {\bibfnamefont {H.}~\bibnamefont {Th{\'e}rien-Aubin}}, \bibinfo {author}
  {\bibfnamefont {S.}~\bibnamefont {Zhu}}, \bibinfo {author} {\bibfnamefont
  {B.}~\bibnamefont {Yang}}, \bibinfo {author} {\bibfnamefont {A.~S.}\
  \bibnamefont {Helmy}}, \bibinfo {author} {\bibfnamefont {O.~D.}\ \bibnamefont
  {Lavrentovich}}, \emph {et~al.},\ }\href@noop {} {\bibfield  {journal}
  {\bibinfo  {journal} {Nature communications}\ }\textbf {\bibinfo {volume}
  {7}},\ \bibinfo {pages} {1} (\bibinfo {year} {2016})}\BibitemShut {NoStop}%
\bibitem [{\citenamefont {Ji}\ \emph {et~al.}(2009)\citenamefont {Ji},
  \citenamefont {Lefort},\ and\ \citenamefont {Morineau}}]{Ji2009}%
  \BibitemOpen
  \bibfield  {author} {\bibinfo {author} {\bibfnamefont {Q.}~\bibnamefont
  {Ji}}, \bibinfo {author} {\bibfnamefont {R.}~\bibnamefont {Lefort}},\ and\
  \bibinfo {author} {\bibfnamefont {D.}~\bibnamefont {Morineau}},\ }\href
  {https://doi.org/https://doi.org/10.1016/j.cplett.2009.07.062} {\bibfield
  {journal} {\bibinfo  {journal} {Chemical Physics Letters}\ }\textbf {\bibinfo
  {volume} {478}},\ \bibinfo {pages} {161} (\bibinfo {year}
  {2009})}\BibitemShut {NoStop}%
\bibitem [{\citenamefont {Madden}\ and\ \citenamefont
  {Glandt}(1988)}]{Madden1988}%
  \BibitemOpen
  \bibfield  {author} {\bibinfo {author} {\bibfnamefont {W.~G.}\ \bibnamefont
  {Madden}}\ and\ \bibinfo {author} {\bibfnamefont {E.~D.}\ \bibnamefont
  {Glandt}},\ }\href@noop {} {\bibfield  {journal} {\bibinfo  {journal}
  {Journal of Statistical Physics}\ }\textbf {\bibinfo {volume} {51}},\
  \bibinfo {pages} {537} (\bibinfo {year} {1988})}\BibitemShut {NoStop}%
\bibitem [{\citenamefont {Given}\ and\ \citenamefont
  {Stell}(1992)}]{given1992}%
  \BibitemOpen
  \bibfield  {author} {\bibinfo {author} {\bibfnamefont {J.~A.}\ \bibnamefont
  {Given}}\ and\ \bibinfo {author} {\bibfnamefont {G.}~\bibnamefont {Stell}},\
  }\href@noop {} {\bibfield  {journal} {\bibinfo  {journal} {The Journal of
  Chemical Physics}\ }\textbf {\bibinfo {volume} {97}},\ \bibinfo {pages}
  {4573} (\bibinfo {year} {1992})}\BibitemShut {NoStop}%
\bibitem [{\citenamefont {Schmidt}(2005)}]{Schmidt2005}%
  \BibitemOpen
  \bibfield  {author} {\bibinfo {author} {\bibfnamefont {M.}~\bibnamefont
  {Schmidt}},\ }\href@noop {} {\bibfield  {journal} {\bibinfo  {journal}
  {Journal of Physics: Condensed Matter}\ }\textbf {\bibinfo {volume} {17}},\
  \bibinfo {pages} {S3481} (\bibinfo {year} {2005})}\BibitemShut {NoStop}%
\bibitem [{\citenamefont {Chan}\ \emph {et~al.}(1988)\citenamefont {Chan},
  \citenamefont {Blum}, \citenamefont {Murphy}, \citenamefont {Wong},\ and\
  \citenamefont {Reppy}}]{Chan1988}%
  \BibitemOpen
  \bibfield  {author} {\bibinfo {author} {\bibfnamefont {M.~H.~W.}\
  \bibnamefont {Chan}}, \bibinfo {author} {\bibfnamefont {K.~I.}\ \bibnamefont
  {Blum}}, \bibinfo {author} {\bibfnamefont {S.~Q.}\ \bibnamefont {Murphy}},
  \bibinfo {author} {\bibfnamefont {G.~K.~S.}\ \bibnamefont {Wong}},\ and\
  \bibinfo {author} {\bibfnamefont {J.~D.}\ \bibnamefont {Reppy}},\ }\href
  {https://doi.org/10.1103/PhysRevLett.61.1950} {\bibfield  {journal} {\bibinfo
   {journal} {Phys. Rev. Lett.}\ }\textbf {\bibinfo {volume} {61}},\ \bibinfo
  {pages} {1950} (\bibinfo {year} {1988})}\BibitemShut {NoStop}%
\bibitem [{\citenamefont {Dadmun}\ and\ \citenamefont
  {Muthukumar}(1993)}]{Dadmun1993}%
  \BibitemOpen
  \bibfield  {author} {\bibinfo {author} {\bibfnamefont {M.~D.}\ \bibnamefont
  {Dadmun}}\ and\ \bibinfo {author} {\bibfnamefont {M.}~\bibnamefont
  {Muthukumar}},\ }\href {http://aip.scitation.org/doi/10.1063/1.464994}
  {\bibfield  {journal} {\bibinfo  {journal} {The Journal of Chemical Physics}\
  }\textbf {\bibinfo {volume} {98}},\ \bibinfo {pages} {4850} (\bibinfo {year}
  {1993})}\BibitemShut {NoStop}%
\bibitem [{\citenamefont {Brown}\ \emph {et~al.}(1998)\citenamefont {Brown},
  \citenamefont {Sokol},\ and\ \citenamefont {Ehrlich}}]{Brown1998}%
  \BibitemOpen
  \bibfield  {author} {\bibinfo {author} {\bibfnamefont {D.~W.}\ \bibnamefont
  {Brown}}, \bibinfo {author} {\bibfnamefont {P.~E.}\ \bibnamefont {Sokol}},\
  and\ \bibinfo {author} {\bibfnamefont {S.~N.}\ \bibnamefont {Ehrlich}},\
  }\href {https://doi.org/10.1103/PhysRevLett.81.1019} {\bibfield  {journal}
  {\bibinfo  {journal} {Phys. Rev. Lett.}\ }\textbf {\bibinfo {volume} {81}},\
  \bibinfo {pages} {1019} (\bibinfo {year} {1998})}\BibitemShut {NoStop}%
\bibitem [{\citenamefont {Iannacchione}\ \emph {et~al.}(1993)\citenamefont
  {Iannacchione}, \citenamefont {Crawford}, \citenamefont
  {\ifmmode~\check{Z}\else \v{Z}\fi{}umer}, \citenamefont {Doane},\ and\
  \citenamefont {Finotello}}]{Iannacchione1993}%
  \BibitemOpen
  \bibfield  {author} {\bibinfo {author} {\bibfnamefont {G.~S.}\ \bibnamefont
  {Iannacchione}}, \bibinfo {author} {\bibfnamefont {G.~P.}\ \bibnamefont
  {Crawford}}, \bibinfo {author} {\bibfnamefont {S.}~\bibnamefont
  {\ifmmode~\check{Z}\else \v{Z}\fi{}umer}}, \bibinfo {author} {\bibfnamefont
  {J.~W.}\ \bibnamefont {Doane}},\ and\ \bibinfo {author} {\bibfnamefont
  {D.}~\bibnamefont {Finotello}},\ }\href
  {https://doi.org/10.1103/PhysRevLett.71.2595} {\bibfield  {journal} {\bibinfo
   {journal} {Phys. Rev. Lett.}\ }\textbf {\bibinfo {volume} {71}},\ \bibinfo
  {pages} {2595} (\bibinfo {year} {1993})}\BibitemShut {NoStop}%
\bibitem [{\citenamefont {Kityk}\ \emph {et~al.}(2008)\citenamefont {Kityk},
  \citenamefont {Wolff}, \citenamefont {Knorr}, \citenamefont {Morineau},
  \citenamefont {Lefort},\ and\ \citenamefont {Huber}}]{Kityk2008}%
  \BibitemOpen
  \bibfield  {author} {\bibinfo {author} {\bibfnamefont {A.~V.}\ \bibnamefont
  {Kityk}}, \bibinfo {author} {\bibfnamefont {M.}~\bibnamefont {Wolff}},
  \bibinfo {author} {\bibfnamefont {K.}~\bibnamefont {Knorr}}, \bibinfo
  {author} {\bibfnamefont {D.}~\bibnamefont {Morineau}}, \bibinfo {author}
  {\bibfnamefont {R.}~\bibnamefont {Lefort}},\ and\ \bibinfo {author}
  {\bibfnamefont {P.}~\bibnamefont {Huber}},\ }\href
  {https://doi.org/10.1103/PhysRevLett.101.187801} {\bibfield  {journal}
  {\bibinfo  {journal} {Phys. Rev. Lett.}\ }\textbf {\bibinfo {volume} {101}},\
  \bibinfo {pages} {187801} (\bibinfo {year} {2008})}\BibitemShut {NoStop}%
\bibitem [{\citenamefont {O\ifmmode~\breve{g}\else \u{g}\fi{}uz}\ \emph
  {et~al.}(2012)\citenamefont {O\ifmmode~\breve{g}\else \u{g}\fi{}uz},
  \citenamefont {Marechal}, \citenamefont {Ramiro-Manzano}, \citenamefont
  {Rodriguez}, \citenamefont {Messina}, \citenamefont {Meseguer},\ and\
  \citenamefont {L\"owen}}]{Erdal2012}%
  \BibitemOpen
  \bibfield  {author} {\bibinfo {author} {\bibfnamefont {E.~C.}\ \bibnamefont
  {O\ifmmode~\breve{g}\else \u{g}\fi{}uz}}, \bibinfo {author} {\bibfnamefont
  {M.}~\bibnamefont {Marechal}}, \bibinfo {author} {\bibfnamefont
  {F.}~\bibnamefont {Ramiro-Manzano}}, \bibinfo {author} {\bibfnamefont
  {I.}~\bibnamefont {Rodriguez}}, \bibinfo {author} {\bibfnamefont
  {R.}~\bibnamefont {Messina}}, \bibinfo {author} {\bibfnamefont {F.~J.}\
  \bibnamefont {Meseguer}},\ and\ \bibinfo {author} {\bibfnamefont
  {H.}~\bibnamefont {L\"owen}},\ }\href
  {https://doi.org/10.1103/PhysRevLett.109.218301} {\bibfield  {journal}
  {\bibinfo  {journal} {Phys. Rev. Lett.}\ }\textbf {\bibinfo {volume} {109}},\
  \bibinfo {pages} {218301} (\bibinfo {year} {2012})}\BibitemShut {NoStop}%
\bibitem [{\citenamefont {Maddox}\ and\ \citenamefont
  {Gubbins}(1997)}]{Maddox1997}%
  \BibitemOpen
  \bibfield  {author} {\bibinfo {author} {\bibfnamefont {M.~W.}\ \bibnamefont
  {Maddox}}\ and\ \bibinfo {author} {\bibfnamefont {K.~E.}\ \bibnamefont
  {Gubbins}},\ }\href {https://doi.org/10.1063/1.475261} {\bibfield  {journal}
  {\bibinfo  {journal} {The Journal of Chemical Physics}\ }\textbf {\bibinfo
  {volume} {107}},\ \bibinfo {pages} {9659} (\bibinfo {year}
  {1997})}\BibitemShut {NoStop}%
\bibitem [{\citenamefont {Gelb}\ \emph {et~al.}(1999)\citenamefont {Gelb},
  \citenamefont {Gubbins}, \citenamefont {Radhakrishnan},\ and\ \citenamefont
  {Sliwinska-Bartkowiak}}]{Gelb1999Review}%
  \BibitemOpen
  \bibfield  {author} {\bibinfo {author} {\bibfnamefont {L.~D.}\ \bibnamefont
  {Gelb}}, \bibinfo {author} {\bibfnamefont {K.}~\bibnamefont {Gubbins}},
  \bibinfo {author} {\bibfnamefont {R.}~\bibnamefont {Radhakrishnan}},\ and\
  \bibinfo {author} {\bibfnamefont {M.}~\bibnamefont {Sliwinska-Bartkowiak}},\
  }\href@noop {} {\bibfield  {journal} {\bibinfo  {journal} {Reports on
  Progress in Physics}\ }\textbf {\bibinfo {volume} {62}},\ \bibinfo {pages}
  {1573} (\bibinfo {year} {1999})}\BibitemShut {NoStop}%
\bibitem [{\citenamefont {Alba-Simionesco}\ \emph {et~al.}(2006)\citenamefont
  {Alba-Simionesco}, \citenamefont {Coasne}, \citenamefont {Dosseh},
  \citenamefont {Dudziak}, \citenamefont {Gubbins}, \citenamefont
  {Radhakrishnan},\ and\ \citenamefont
  {Sliwinska-Bartkowiak}}]{Alba2006Review}%
  \BibitemOpen
  \bibfield  {author} {\bibinfo {author} {\bibfnamefont {C.}~\bibnamefont
  {Alba-Simionesco}}, \bibinfo {author} {\bibfnamefont {B.}~\bibnamefont
  {Coasne}}, \bibinfo {author} {\bibfnamefont {G.}~\bibnamefont {Dosseh}},
  \bibinfo {author} {\bibfnamefont {G.}~\bibnamefont {Dudziak}}, \bibinfo
  {author} {\bibfnamefont {K.}~\bibnamefont {Gubbins}}, \bibinfo {author}
  {\bibfnamefont {R.}~\bibnamefont {Radhakrishnan}},\ and\ \bibinfo {author}
  {\bibfnamefont {M.}~\bibnamefont {Sliwinska-Bartkowiak}},\ }\href@noop {}
  {\bibfield  {journal} {\bibinfo  {journal} {Journal of Physics: Condensed
  Matter}\ }\textbf {\bibinfo {volume} {18}},\ \bibinfo {pages} {R15} (\bibinfo
  {year} {2006})}\BibitemShut {NoStop}%
\bibitem [{\citenamefont {de~las Heras}\ \emph {et~al.}(2005)\citenamefont
  {de~las Heras}, \citenamefont {Velasco},\ and\ \citenamefont
  {Mederos}}]{delasHeras2005}%
  \BibitemOpen
  \bibfield  {author} {\bibinfo {author} {\bibfnamefont {D.}~\bibnamefont
  {de~las Heras}}, \bibinfo {author} {\bibfnamefont {E.}~\bibnamefont
  {Velasco}},\ and\ \bibinfo {author} {\bibfnamefont {L.}~\bibnamefont
  {Mederos}},\ }\href {https://doi.org/10.1103/PhysRevLett.94.017801}
  {\bibfield  {journal} {\bibinfo  {journal} {Phys. Rev. Lett.}\ }\textbf
  {\bibinfo {volume} {94}},\ \bibinfo {pages} {017801} (\bibinfo {year}
  {2005})}\BibitemShut {NoStop}%
\bibitem [{\citenamefont {Gefen}\ \emph {et~al.}(1980)\citenamefont {Gefen},
  \citenamefont {Mandelbrot},\ and\ \citenamefont {Aharony}}]{Gefen1980}%
  \BibitemOpen
  \bibfield  {author} {\bibinfo {author} {\bibfnamefont {Y.}~\bibnamefont
  {Gefen}}, \bibinfo {author} {\bibfnamefont {B.~B.}\ \bibnamefont
  {Mandelbrot}},\ and\ \bibinfo {author} {\bibfnamefont {A.}~\bibnamefont
  {Aharony}},\ }\href {https://doi.org/10.1103/PhysRevLett.45.855} {\bibfield
  {journal} {\bibinfo  {journal} {Phys. Rev. Lett.}\ }\textbf {\bibinfo
  {volume} {45}},\ \bibinfo {pages} {855} (\bibinfo {year} {1980})}\BibitemShut
  {NoStop}%
\bibitem [{\citenamefont {Gefen}\ \emph {et~al.}(1983)\citenamefont {Gefen},
  \citenamefont {Aharony},\ and\ \citenamefont {Mandelbrot}}]{Gefen1983}%
  \BibitemOpen
  \bibfield  {author} {\bibinfo {author} {\bibfnamefont {Y.}~\bibnamefont
  {Gefen}}, \bibinfo {author} {\bibfnamefont {A.}~\bibnamefont {Aharony}},\
  and\ \bibinfo {author} {\bibfnamefont {B.~B.}\ \bibnamefont {Mandelbrot}},\
  }\href {https://doi.org/10.1088/0305-4470/16/6/021} {\bibfield  {journal}
  {\bibinfo  {journal} {Journal of Physics A: Mathematical and General}\
  }\textbf {\bibinfo {volume} {16}},\ \bibinfo {pages} {1267} (\bibinfo {year}
  {1983})}\BibitemShut {NoStop}%
\bibitem [{\citenamefont {Gefen}\ \emph
  {et~al.}(1984{\natexlab{a}})\citenamefont {Gefen}, \citenamefont {Aharony},
  \citenamefont {Shapir},\ and\ \citenamefont {Mandelbrot}}]{Gefen1984a}%
  \BibitemOpen
  \bibfield  {author} {\bibinfo {author} {\bibfnamefont {Y.}~\bibnamefont
  {Gefen}}, \bibinfo {author} {\bibfnamefont {A.}~\bibnamefont {Aharony}},
  \bibinfo {author} {\bibfnamefont {Y.}~\bibnamefont {Shapir}},\ and\ \bibinfo
  {author} {\bibfnamefont {B.~B.}\ \bibnamefont {Mandelbrot}},\ }\href
  {https://doi.org/10.1088/0305-4470/17/2/028} {\bibfield  {journal} {\bibinfo
  {journal} {Journal of Physics A: Mathematical and General}\ }\textbf
  {\bibinfo {volume} {17}},\ \bibinfo {pages} {435} (\bibinfo {year}
  {1984}{\natexlab{a}})}\BibitemShut {NoStop}%
\bibitem [{\citenamefont {Gefen}\ \emph
  {et~al.}(1984{\natexlab{b}})\citenamefont {Gefen}, \citenamefont {Aharony},\
  and\ \citenamefont {Mandelbrot}}]{Gefen1984}%
  \BibitemOpen
  \bibfield  {author} {\bibinfo {author} {\bibfnamefont {Y.}~\bibnamefont
  {Gefen}}, \bibinfo {author} {\bibfnamefont {A.}~\bibnamefont {Aharony}},\
  and\ \bibinfo {author} {\bibfnamefont {B.~B.}\ \bibnamefont {Mandelbrot}},\
  }\href {https://doi.org/10.1088/0305-4470/17/6/024} {\bibfield  {journal}
  {\bibinfo  {journal} {Journal of Physics A: Mathematical and General}\
  }\textbf {\bibinfo {volume} {17}},\ \bibinfo {pages} {1277} (\bibinfo {year}
  {1984}{\natexlab{b}})}\BibitemShut {NoStop}%
\bibitem [{\citenamefont {Windus}\ and\ \citenamefont
  {Jensen}(2009)}]{Windus2009}%
  \BibitemOpen
  \bibfield  {author} {\bibinfo {author} {\bibfnamefont {A.~L.}\ \bibnamefont
  {Windus}}\ and\ \bibinfo {author} {\bibfnamefont {H.~J.}\ \bibnamefont
  {Jensen}},\ }\href {https://doi.org/10.1016/j.physa.2009.04.008} {\bibfield
  {journal} {\bibinfo  {journal} {Physica A: Statistical Mechanics and its
  Applications}\ }\textbf {\bibinfo {volume} {388}},\ \bibinfo {pages} {3107}
  (\bibinfo {year} {2009})}\BibitemShut {NoStop}%
\bibitem [{\citenamefont {Popa-Nita}\ and\ \citenamefont
  {Romano}(2001)}]{Popa-Nita2001}%
  \BibitemOpen
  \bibfield  {author} {\bibinfo {author} {\bibfnamefont {V.}~\bibnamefont
  {Popa-Nita}}\ and\ \bibinfo {author} {\bibfnamefont {S.}~\bibnamefont
  {Romano}},\ }\href {https://doi.org/10.1016/S0301-0104(00)00340-2} {\bibfield
   {journal} {\bibinfo  {journal} {Chemical Physics}\ }\textbf {\bibinfo
  {volume} {264}},\ \bibinfo {pages} {91} (\bibinfo {year} {2001})}\BibitemShut
  {NoStop}%
\bibitem [{\citenamefont {Radzihovsky}\ and\ \citenamefont
  {Toner}(1999)}]{Radzihovsky1999}%
  \BibitemOpen
  \bibfield  {author} {\bibinfo {author} {\bibfnamefont {L.}~\bibnamefont
  {Radzihovsky}}\ and\ \bibinfo {author} {\bibfnamefont {J.}~\bibnamefont
  {Toner}},\ }\href {https://doi.org/10.1103/PhysRevB.60.206} {\bibfield
  {journal} {\bibinfo  {journal} {Phys. Rev. B}\ }\textbf {\bibinfo {volume}
  {60}},\ \bibinfo {pages} {206} (\bibinfo {year} {1999})}\BibitemShut
  {NoStop}%
\bibitem [{\citenamefont {Feldman}(2000)}]{Feldman2000}%
  \BibitemOpen
  \bibfield  {author} {\bibinfo {author} {\bibfnamefont {D.~E.}\ \bibnamefont
  {Feldman}},\ }\href {https://doi.org/10.1103/PhysRevLett.84.4886} {\bibfield
  {journal} {\bibinfo  {journal} {Phys. Rev. Lett.}\ }\textbf {\bibinfo
  {volume} {84}},\ \bibinfo {pages} {4886} (\bibinfo {year}
  {2000})}\BibitemShut {NoStop}%
\bibitem [{\citenamefont {Hvozd}\ \emph {et~al.}(2018)\citenamefont {Hvozd},
  \citenamefont {Patsahan},\ and\ \citenamefont {Holovko}}]{Hvozd2018}%
  \BibitemOpen
  \bibfield  {author} {\bibinfo {author} {\bibfnamefont {M.}~\bibnamefont
  {Hvozd}}, \bibinfo {author} {\bibfnamefont {T.}~\bibnamefont {Patsahan}},\
  and\ \bibinfo {author} {\bibfnamefont {M.}~\bibnamefont {Holovko}},\ }\href
  {https://doi.org/10.1021/acs.jpcb.7b11834} {\bibfield  {journal} {\bibinfo
  {journal} {The Journal of Physical Chemistry B}\ }\textbf {\bibinfo {volume}
  {122}},\ \bibinfo {pages} {5534} (\bibinfo {year} {2018})}\BibitemShut
  {NoStop}%
\bibitem [{\citenamefont {Torquato}(2002)}]{Torquato2002}%
  \BibitemOpen
  \bibfield  {author} {\bibinfo {author} {\bibfnamefont {S.}~\bibnamefont
  {Torquato}},\ }\href {https://doi.org/10.1007/978-1-4757-6355-3} {\emph
  {\bibinfo {title} {{Random Heterogeneous Materials: Microstructure and
  Macroscopic Properties}}}},\ Vol.~\bibinfo {volume} {16}\ (\bibinfo
  {publisher} {Springer},\ \bibinfo {address} {New York},\ \bibinfo {year}
  {2002})\BibitemShut {NoStop}%
\bibitem [{\citenamefont {DeSimone}\ \emph {et~al.}(1986)\citenamefont
  {DeSimone}, \citenamefont {Demoulini},\ and\ \citenamefont
  {Stratt}}]{DeSimone1986}%
  \BibitemOpen
  \bibfield  {author} {\bibinfo {author} {\bibfnamefont {T.}~\bibnamefont
  {DeSimone}}, \bibinfo {author} {\bibfnamefont {S.}~\bibnamefont
  {Demoulini}},\ and\ \bibinfo {author} {\bibfnamefont {R.~M.}\ \bibnamefont
  {Stratt}},\ }\href {https://doi.org/10.1063/1.451615} {\bibfield  {journal}
  {\bibinfo  {journal} {The Journal of Chemical Physics}\ }\textbf {\bibinfo
  {volume} {85}},\ \bibinfo {pages} {391} (\bibinfo {year} {1986})}\BibitemShut
  {NoStop}%
\bibitem [{\citenamefont {Drory}(1997)}]{Drory1997}%
  \BibitemOpen
  \bibfield  {author} {\bibinfo {author} {\bibfnamefont {A.}~\bibnamefont
  {Drory}},\ }\href {https://doi.org/10.1103/PhysRevE.55.3878} {\bibfield
  {journal} {\bibinfo  {journal} {Physical Review E}\ }\textbf {\bibinfo
  {volume} {55}},\ \bibinfo {pages} {3878} (\bibinfo {year}
  {1997})}\BibitemShut {NoStop}%
\bibitem [{\citenamefont {Lorenz}\ and\ \citenamefont
  {Ziff}(2001)}]{Lorenz2001a}%
  \BibitemOpen
  \bibfield  {author} {\bibinfo {author} {\bibfnamefont {C.~D.}\ \bibnamefont
  {Lorenz}}\ and\ \bibinfo {author} {\bibfnamefont {R.~M.}\ \bibnamefont
  {Ziff}},\ }\href {https://doi.org/10.1063/1.1338506} {\bibfield  {journal}
  {\bibinfo  {journal} {The Journal of Chemical Physics}\ }\textbf {\bibinfo
  {volume} {114}},\ \bibinfo {pages} {3659} (\bibinfo {year}
  {2001})}\BibitemShut {NoStop}%
\bibitem [{\citenamefont {Torquato}\ and\ \citenamefont
  {Jiao}(2012)}]{TorquatoJiao2012}%
  \BibitemOpen
  \bibfield  {author} {\bibinfo {author} {\bibfnamefont {S.}~\bibnamefont
  {Torquato}}\ and\ \bibinfo {author} {\bibfnamefont {Y.}~\bibnamefont
  {Jiao}},\ }\href {https://doi.org/10.1063/1.4742750} {\bibfield  {journal}
  {\bibinfo  {journal} {The Journal of Chemical Physics}\ }\textbf {\bibinfo
  {volume} {137}},\ \bibinfo {pages} {074106} (\bibinfo {year}
  {2012})}\BibitemShut {NoStop}%
\bibitem [{SI()}]{SI}%
  \BibitemOpen
  \href@noop {} {}\bibinfo {note} {See Supplemental Material for a discussion
  of our numerical approach, the influence of $\sigma/\lambda$ on the critical
  exponent $\gamma$ and the analytical expression for the mean cluster size in
  $D = 5$.}\BibitemShut {Stop}%
\bibitem [{\citenamefont {Balberg}\ \emph {et~al.}(1984)\citenamefont
  {Balberg}, \citenamefont {Anderson}, \citenamefont {Alexander},\ and\
  \citenamefont {Wagner}}]{Balberg1984}%
  \BibitemOpen
  \bibfield  {author} {\bibinfo {author} {\bibfnamefont {I.}~\bibnamefont
  {Balberg}}, \bibinfo {author} {\bibfnamefont {C.~H.}\ \bibnamefont
  {Anderson}}, \bibinfo {author} {\bibfnamefont {S.}~\bibnamefont
  {Alexander}},\ and\ \bibinfo {author} {\bibfnamefont {N.}~\bibnamefont
  {Wagner}},\ }\href {https://doi.org/10.1103/PhysRevB.30.3933} {\bibfield
  {journal} {\bibinfo  {journal} {Phys. Rev. B}\ }\textbf {\bibinfo {volume}
  {30}},\ \bibinfo {pages} {3933} (\bibinfo {year} {1984})}\BibitemShut
  {NoStop}%
\bibitem [{\citenamefont {Torquato}(2012)}]{Torquato2012}%
  \BibitemOpen
  \bibfield  {author} {\bibinfo {author} {\bibfnamefont {S.}~\bibnamefont
  {Torquato}},\ }\href {https://doi.org/10.1063/1.3679861} {\bibfield
  {journal} {\bibinfo  {journal} {The Journal of Chemical Physics}\ }\textbf
  {\bibinfo {volume} {136}},\ \bibinfo {pages} {054106} (\bibinfo {year}
  {2012})}\BibitemShut {NoStop}%
\bibitem [{\citenamefont {Grimaldi}(2015)}]{Grimaldi2015}%
  \BibitemOpen
  \bibfield  {author} {\bibinfo {author} {\bibfnamefont {C.}~\bibnamefont
  {Grimaldi}},\ }\href@noop {} {\bibfield  {journal} {\bibinfo  {journal}
  {Physical Review E}\ }\textbf {\bibinfo {volume} {92}},\ \bibinfo {pages}
  {012126} (\bibinfo {year} {2015})}\BibitemShut {NoStop}%
\bibitem [{\citenamefont {Lee}\ and\ \citenamefont {Torquato}(1990)}]{Lee1990}%
  \BibitemOpen
  \bibfield  {author} {\bibinfo {author} {\bibfnamefont {S.~B.}\ \bibnamefont
  {Lee}}\ and\ \bibinfo {author} {\bibfnamefont {S.}~\bibnamefont {Torquato}},\
  }\href {https://doi.org/10.1103/PhysRevA.41.5338} {\bibfield  {journal}
  {\bibinfo  {journal} {Physical Review A}\ }\textbf {\bibinfo {volume} {41}},\
  \bibinfo {pages} {5338} (\bibinfo {year} {1990})}\BibitemShut {NoStop}%
\bibitem [{\citenamefont {Miller}(2009)}]{Miller2009}%
  \BibitemOpen
  \bibfield  {author} {\bibinfo {author} {\bibfnamefont {M.~A.}\ \bibnamefont
  {Miller}},\ }\href {https://doi.org/10.1063/1.3204483} {\bibfield  {journal}
  {\bibinfo  {journal} {The Journal of Chemical Physics}\ }\textbf {\bibinfo
  {volume} {131}},\ \bibinfo {pages} {066101} (\bibinfo {year}
  {2009})}\BibitemShut {NoStop}%
\bibitem [{\citenamefont {Coniglio}\ \emph {et~al.}(1977)\citenamefont
  {Coniglio}, \citenamefont {Angelis},\ and\ \citenamefont
  {Forlani}}]{Coniglio1977}%
  \BibitemOpen
  \bibfield  {author} {\bibinfo {author} {\bibfnamefont {A.}~\bibnamefont
  {Coniglio}}, \bibinfo {author} {\bibfnamefont {U.~D.}\ \bibnamefont
  {Angelis}},\ and\ \bibinfo {author} {\bibfnamefont {A.}~\bibnamefont
  {Forlani}},\ }\href {https://doi.org/10.1088/0305-4470/10/7/011} {\bibfield
  {journal} {\bibinfo  {journal} {Journal of Physics A: Mathematical and
  General}\ }\textbf {\bibinfo {volume} {10}},\ \bibinfo {pages} {1123}
  (\bibinfo {year} {1977})}\BibitemShut {NoStop}%
\bibitem [{\citenamefont {Hansen}\ and\ \citenamefont
  {McDonald}(2013)}]{Hansen2013}%
  \BibitemOpen
  \bibfield  {author} {\bibinfo {author} {\bibfnamefont {J.~P.}\ \bibnamefont
  {Hansen}}\ and\ \bibinfo {author} {\bibfnamefont {I.~R.}\ \bibnamefont
  {McDonald}},\ }\href {https://doi.org/10.1016/C2010-0-66723-X} {\emph
  {\bibinfo {title} {Theory of Simple Liquids: With Applications to Soft
  Matter: Fourth Edition}}}\ (\bibinfo  {publisher} {Academic Press},\ \bibinfo
  {address} {Oxford},\ \bibinfo {year} {2013})\BibitemShut {NoStop}%
\bibitem [{\citenamefont {Bug}\ \emph {et~al.}(1985)\citenamefont {Bug},
  \citenamefont {Safran}, \citenamefont {Grest},\ and\ \citenamefont
  {Webman}}]{Bug1985}%
  \BibitemOpen
  \bibfield  {author} {\bibinfo {author} {\bibfnamefont {A.~L.~R.}\
  \bibnamefont {Bug}}, \bibinfo {author} {\bibfnamefont {S.~A.}\ \bibnamefont
  {Safran}}, \bibinfo {author} {\bibfnamefont {G.~S.}\ \bibnamefont {Grest}},\
  and\ \bibinfo {author} {\bibfnamefont {I.}~\bibnamefont {Webman}},\ }\href
  {https://doi.org/10.1103/PhysRevLett.55.1896} {\bibfield  {journal} {\bibinfo
   {journal} {Phys. Rev. Lett.}\ }\textbf {\bibinfo {volume} {55}},\ \bibinfo
  {pages} {1896} (\bibinfo {year} {1985})}\BibitemShut {NoStop}%
\bibitem [{\citenamefont {Grimaldi}(2017)}]{Grimaldi2017b}%
  \BibitemOpen
  \bibfield  {author} {\bibinfo {author} {\bibfnamefont {C.}~\bibnamefont
  {Grimaldi}},\ }\href {https://doi.org/10.1063/1.4991093} {\bibfield
  {journal} {\bibinfo  {journal} {The Journal of Chemical Physics}\ }\textbf
  {\bibinfo {volume} {147}},\ \bibinfo {pages} {074502} (\bibinfo {year}
  {2017})}\BibitemShut {NoStop}%
\bibitem [{\citenamefont {Chiew}\ and\ \citenamefont
  {Glandt}(1983)}]{Chiew1983}%
  \BibitemOpen
  \bibfield  {author} {\bibinfo {author} {\bibfnamefont {Y.~C.}\ \bibnamefont
  {Chiew}}\ and\ \bibinfo {author} {\bibfnamefont {E.~D.}\ \bibnamefont
  {Glandt}},\ }\href {https://doi.org/10.1088/0305-4470/16/11/026} {\bibfield
  {journal} {\bibinfo  {journal} {Journal of Physics A: Mathematical and
  General}\ }\textbf {\bibinfo {volume} {16}},\ \bibinfo {pages} {2599}
  (\bibinfo {year} {1983})}\BibitemShut {NoStop}%
\bibitem [{\citenamefont {Baxter}(1982)}]{Baxter1982}%
  \BibitemOpen
  \bibfield  {author} {\bibinfo {author} {\bibfnamefont {R.~J.}\ \bibnamefont
  {Baxter}},\ }\href@noop {} {\emph {\bibinfo {title} {Exactly solved models in
  statistical mechanics}}}\ (\bibinfo  {publisher} {Academic Press Limited},\
  \bibinfo {address} {London},\ \bibinfo {year} {1982})\BibitemShut {NoStop}%
\bibitem [{\citenamefont {Adler}\ \emph {et~al.}(1990)\citenamefont {Adler},
  \citenamefont {Meir}, \citenamefont {Aharony},\ and\ \citenamefont
  {Harris}}]{Adler1990}%
  \BibitemOpen
  \bibfield  {author} {\bibinfo {author} {\bibfnamefont {J.}~\bibnamefont
  {Adler}}, \bibinfo {author} {\bibfnamefont {Y.}~\bibnamefont {Meir}},
  \bibinfo {author} {\bibfnamefont {A.}~\bibnamefont {Aharony}},\ and\ \bibinfo
  {author} {\bibfnamefont {A.~B.}\ \bibnamefont {Harris}},\ }\href
  {https://doi.org/10.1103/PhysRevB.41.9183} {\bibfield  {journal} {\bibinfo
  {journal} {Physical Review B}\ }\textbf {\bibinfo {volume} {41}},\ \bibinfo
  {pages} {9183} (\bibinfo {year} {1990})}\BibitemShut {NoStop}%
\bibitem [{\citenamefont {Coupette}\ \emph {et~al.}(2021)\citenamefont
  {Coupette}, \citenamefont {de~Bruijn}, \citenamefont {Bult}, \citenamefont
  {Finner}, \citenamefont {Miller}, \citenamefont {van~der Schoot},\ and\
  \citenamefont {Schilling}}]{Coupette2021}%
  \BibitemOpen
  \bibfield  {author} {\bibinfo {author} {\bibfnamefont {F.}~\bibnamefont
  {Coupette}}, \bibinfo {author} {\bibfnamefont {R.}~\bibnamefont {de~Bruijn}},
  \bibinfo {author} {\bibfnamefont {P.}~\bibnamefont {Bult}}, \bibinfo {author}
  {\bibfnamefont {S.}~\bibnamefont {Finner}}, \bibinfo {author} {\bibfnamefont
  {M.~A.}\ \bibnamefont {Miller}}, \bibinfo {author} {\bibfnamefont
  {P.}~\bibnamefont {van~der Schoot}},\ and\ \bibinfo {author} {\bibfnamefont
  {T.}~\bibnamefont {Schilling}},\ }\href
  {https://doi.org/10.1103/PhysRevE.103.042115} {\bibfield  {journal} {\bibinfo
   {journal} {Phys. Rev. E}\ }\textbf {\bibinfo {volume} {103}},\ \bibinfo
  {pages} {042115} (\bibinfo {year} {2021})}\BibitemShut {NoStop}%
\bibitem [{\citenamefont {Coupette}\ \emph {et~al.}(2020)\citenamefont
  {Coupette}, \citenamefont {H{\"a}rtel},\ and\ \citenamefont
  {Schilling}}]{Coupette2020}%
  \BibitemOpen
  \bibfield  {author} {\bibinfo {author} {\bibfnamefont {F.}~\bibnamefont
  {Coupette}}, \bibinfo {author} {\bibfnamefont {A.}~\bibnamefont
  {H{\"a}rtel}},\ and\ \bibinfo {author} {\bibfnamefont {T.}~\bibnamefont
  {Schilling}},\ }\href@noop {} {\bibfield  {journal} {\bibinfo  {journal}
  {Physical Review E}\ }\textbf {\bibinfo {volume} {101}},\ \bibinfo {pages}
  {062126} (\bibinfo {year} {2020})}\BibitemShut {NoStop}%
\end{thebibliography}%


\begin{thebibliography}{7}%
\makeatletter
\providecommand \@ifxundefined [1]{%
 \@ifx{#1\undefined}
}%
\providecommand \@ifnum [1]{%
 \ifnum #1\expandafter \@firstoftwo
 \else \expandafter \@secondoftwo
 \fi
}%
\providecommand \@ifx [1]{%
 \ifx #1\expandafter \@firstoftwo
 \else \expandafter \@secondoftwo
 \fi
}%
\providecommand \natexlab [1]{#1}%
\providecommand \enquote  [1]{``#1''}%
\providecommand \bibnamefont  [1]{#1}%
\providecommand \bibfnamefont [1]{#1}%
\providecommand \citenamefont [1]{#1}%
\providecommand \href@noop [0]{\@secondoftwo}%
\providecommand \href [0]{\begingroup \@sanitize@url \@href}%
\providecommand \@href[1]{\@@startlink{#1}\@@href}%
\providecommand \@@href[1]{\endgroup#1\@@endlink}%
\providecommand \@sanitize@url [0]{\catcode `\\12\catcode `\$12\catcode
  `\&12\catcode `\#12\catcode `\^12\catcode `\_12\catcode `\%12\relax}%
\providecommand \@@startlink[1]{}%
\providecommand \@@endlink[0]{}%
\providecommand \url  [0]{\begingroup\@sanitize@url \@url }%
\providecommand \@url [1]{\endgroup\@href {#1}{\urlprefix }}%
\providecommand \urlprefix  [0]{URL }%
\providecommand \Eprint [0]{\href }%
\providecommand \doibase [0]{https://doi.org/}%
\providecommand \selectlanguage [0]{\@gobble}%
\providecommand \bibinfo  [0]{\@secondoftwo}%
\providecommand \bibfield  [0]{\@secondoftwo}%
\providecommand \translation [1]{[#1]}%
\providecommand \BibitemOpen [0]{}%
\providecommand \bibitemStop [0]{}%
\providecommand \bibitemNoStop [0]{.\EOS\space}%
\providecommand \EOS [0]{\spacefactor3000\relax}%
\providecommand \BibitemShut  [1]{\csname bibitem#1\endcsname}%
\let\auto@bib@innerbib\@empty
\bibitem [{\citenamefont {Talman}(1978)}]{Talman1978}%
  \BibitemOpen
  \bibfield  {author} {\bibinfo {author} {\bibfnamefont {J.~D.}\ \bibnamefont
  {Talman}},\ }\href {https://doi.org/10.1016/0021-9991(78)90107-9} {\bibfield
  {journal} {\bibinfo  {journal} {Journal of Computational Physics}\ }\textbf
  {\bibinfo {volume} {29}},\ \bibinfo {pages} {35} (\bibinfo {year}
  {1978})}\BibitemShut {NoStop}%
\bibitem [{\citenamefont {Heinen}\ \emph {et~al.}(2015)\citenamefont {Heinen},
  \citenamefont {Schnyder}, \citenamefont {Brady},\ and\ \citenamefont
  {L{\"{o}}wen}}]{Heinen2015}%
  \BibitemOpen
  \bibfield  {author} {\bibinfo {author} {\bibfnamefont {M.}~\bibnamefont
  {Heinen}}, \bibinfo {author} {\bibfnamefont {S.~K.}\ \bibnamefont
  {Schnyder}}, \bibinfo {author} {\bibfnamefont {J.~F.}\ \bibnamefont
  {Brady}},\ and\ \bibinfo {author} {\bibfnamefont {H.}~\bibnamefont
  {L{\"{o}}wen}},\ }\href {https://doi.org/10.1103/PhysRevLett.115.097801}
  {\bibfield  {journal} {\bibinfo  {journal} {Physical Review Letters}\
  }\textbf {\bibinfo {volume} {115}},\ \bibinfo {pages} {097801} (\bibinfo
  {year} {2015})}\BibitemShut {NoStop}%
\bibitem [{\citenamefont {Hamilton}(2000)}]{Hamilton2000}%
  \BibitemOpen
  \bibfield  {author} {\bibinfo {author} {\bibfnamefont {A.~J.~S.}\
  \bibnamefont {Hamilton}},\ }\href
  {https://doi.org/10.1046/j.1365-8711.2000.03071.x} {\bibfield  {journal}
  {\bibinfo  {journal} {Monthly Notices of the Royal Astronomical Society}\
  }\textbf {\bibinfo {volume} {312}},\ \bibinfo {pages} {257} (\bibinfo {year}
  {2000})}\BibitemShut {NoStop}%
\bibitem [{\citenamefont {Heinen}\ \emph {et~al.}(2014)\citenamefont {Heinen},
  \citenamefont {Allahyarov},\ and\ \citenamefont {L{\"{o}}wen}}]{Heinen2014}%
  \BibitemOpen
  \bibfield  {author} {\bibinfo {author} {\bibfnamefont {M.}~\bibnamefont
  {Heinen}}, \bibinfo {author} {\bibfnamefont {E.}~\bibnamefont {Allahyarov}},\
  and\ \bibinfo {author} {\bibfnamefont {H.}~\bibnamefont {L{\"{o}}wen}},\
  }\href {https://doi.org/10.1002/jcc.23446} {\bibfield  {journal} {\bibinfo
  {journal} {Journal of Computational Chemistry}\ }\textbf {\bibinfo {volume}
  {35}},\ \bibinfo {pages} {275} (\bibinfo {year} {2014})}\BibitemShut
  {NoStop}%
\bibitem [{\citenamefont {Kovalenko}\ \emph {et~al.}(1999)\citenamefont
  {Kovalenko}, \citenamefont {Ten-no},\ and\ \citenamefont
  {Hirata}}]{Kovalenko1999}%
  \BibitemOpen
  \bibfield  {author} {\bibinfo {author} {\bibfnamefont {A.}~\bibnamefont
  {Kovalenko}}, \bibinfo {author} {\bibfnamefont {S.}~\bibnamefont {Ten-no}},\
  and\ \bibinfo {author} {\bibfnamefont {F.}~\bibnamefont {Hirata}},\ }\href
  {https://doi.org/10.1002/(SICI)1096-987X(19990715)20:9<928::AID-JCC4>3.0.CO;2-X}
  {\bibfield  {journal} {\bibinfo  {journal} {Journal of Computational
  Chemistry}\ }\textbf {\bibinfo {volume} {20}},\ \bibinfo {pages} {928}
  (\bibinfo {year} {1999})}\BibitemShut {NoStop}%
\bibitem [{\citenamefont {Stell}(1984)}]{Stell1984a}%
  \BibitemOpen
  \bibfield  {author} {\bibinfo {author} {\bibfnamefont {G.}~\bibnamefont
  {Stell}},\ }\href {https://doi.org/10.1088/0305-4470/17/15/007} {\bibfield
  {journal} {\bibinfo  {journal} {Journal of Physics A: Mathematical and
  General}\ }\textbf {\bibinfo {volume} {17}},\ \bibinfo {pages} {L855}
  (\bibinfo {year} {1984})}\BibitemShut {NoStop}%
\bibitem [{\citenamefont {Leutheusser}(1984)}]{Leutheusser1984}%
  \BibitemOpen
  \bibfield  {author} {\bibinfo {author} {\bibfnamefont {E.}~\bibnamefont
  {Leutheusser}},\ }\href {https://doi.org/10.1016/0378-4371(84)90050-5}
  {\bibfield  {journal} {\bibinfo  {journal} {Physica A: Statistical Mechanics
  and its Applications}\ }\textbf {\bibinfo {volume} {127}},\ \bibinfo {pages}
  {667} (\bibinfo {year} {1984})}\BibitemShut {NoStop}%
\end{thebibliography}%
\end{document}


\title{Supplemental Material for `Connectedness percolation of fractal liquids'}
	\author{René de Bruijn}
	\email{r.a.j.d.bruijn@tue.nl}
	\affiliation{Department of Applied Physics, Eindhoven University of Technology, P.O. Box 513, 5600 MB Eindhoven, Netherlands}
	\author{Paul van der Schoot}
	\affiliation{Department of Applied Physics, Eindhoven University of Technology, P.O. Box 513, 5600 MB Eindhoven, Netherlands}
\maketitle

\section*{Numerical solution strategy}
We determine the onset of the material-spanning cluster, \textit{i.e.}, the percolation threshold, starting from the connectedness Ornstein-Zernike (cOZ) Equation
\begin{equation}\label{eq:Ornstein-ZernikePT}
P(r) = C^+(r) + \rho\int \mathrm{d}^{D}\mathbf{r}'P(r') C^+(|\mathbf{r}-\mathbf{r}'|),
\end{equation}
where $P(r)$ is the pair connectedness function, $C^+(r)$ the direct connectedness function, $\rho$ the number density and $D$ the (spreading) dimension. Moreover, we require the liquid-state Ornstein-Zernike (OZ) Equation to order to obtain the radial distribution function $g(r)$
\begin{equation}
g(r) = 1 + c(r) + \rho\int \mathrm{d}^{D}\textbf{r}'\left[g(|\textbf{r}|)-1\right] c(|\textbf{r} - \textbf{r}'|),
\end{equation} 
where $c(r)$ the direct correlation function.
For isotropic systems these equations reduce to algebraic equations in Fourier-space
\begin{equation}
\widehat{P}(q) = \frac{\widehat{C}^+(q)}{1-\rho \widehat{C}^+(q)},
\end{equation}
and 
\begin{equation}
\widehat{g}(q) = \frac{1 + \widehat{c}(q)}{1-\rho \widehat{c}(q)}.
\end{equation}

Since the (connectedness) Percus-Yevick  closure is handled in real-space, our (iterative) numerical solution strategy employs a spectral solver which is based on the generalized $D$-dimensional Hankel transform pair
\begin{align}
F(q) &= \frac{(2 \pi)^{D/2}}{q^{D/2-1}} \int^\infty_0 \mathrm{d} r r^{D/2}  J_{D/2-1}(q r) f(r), \label{eq:FT} \\
f(r) &= \frac{r^{1-D/2}}{(2 \pi)^{D/2}} \int^\infty_0 \mathrm{d} q q^{D/2}  J_{D/2-1}(q r) F(q), \label{eq:IFT} 
\end{align}
valid for a $D$-dimensional isotropic function $f(r)$. Here, $J_{D/2-1}(x)$ is the Bessel function of the first kind and order $D/2-1$, which is analytic with respect of both $D\geq 1$ and $q, r> 0$. The Hankel transforms are tackled using a sampling technique based on a logarithmic grid \cite{Talman1978,Heinen2015,Hamilton2000, Heinen2014}, which is equivalent to the approach by Heinen and coworkers \cite{Heinen2015}. We solve the set of equations iteratively by a modified Picard iteration, the stability and convergence of which is increased by the modified direct inversion of iterative subspace (MDIIS) approach \cite{Kovalenko1999}. 

\begin{figure}[bt]
	\centering
	\includegraphics[width=8.6cm]{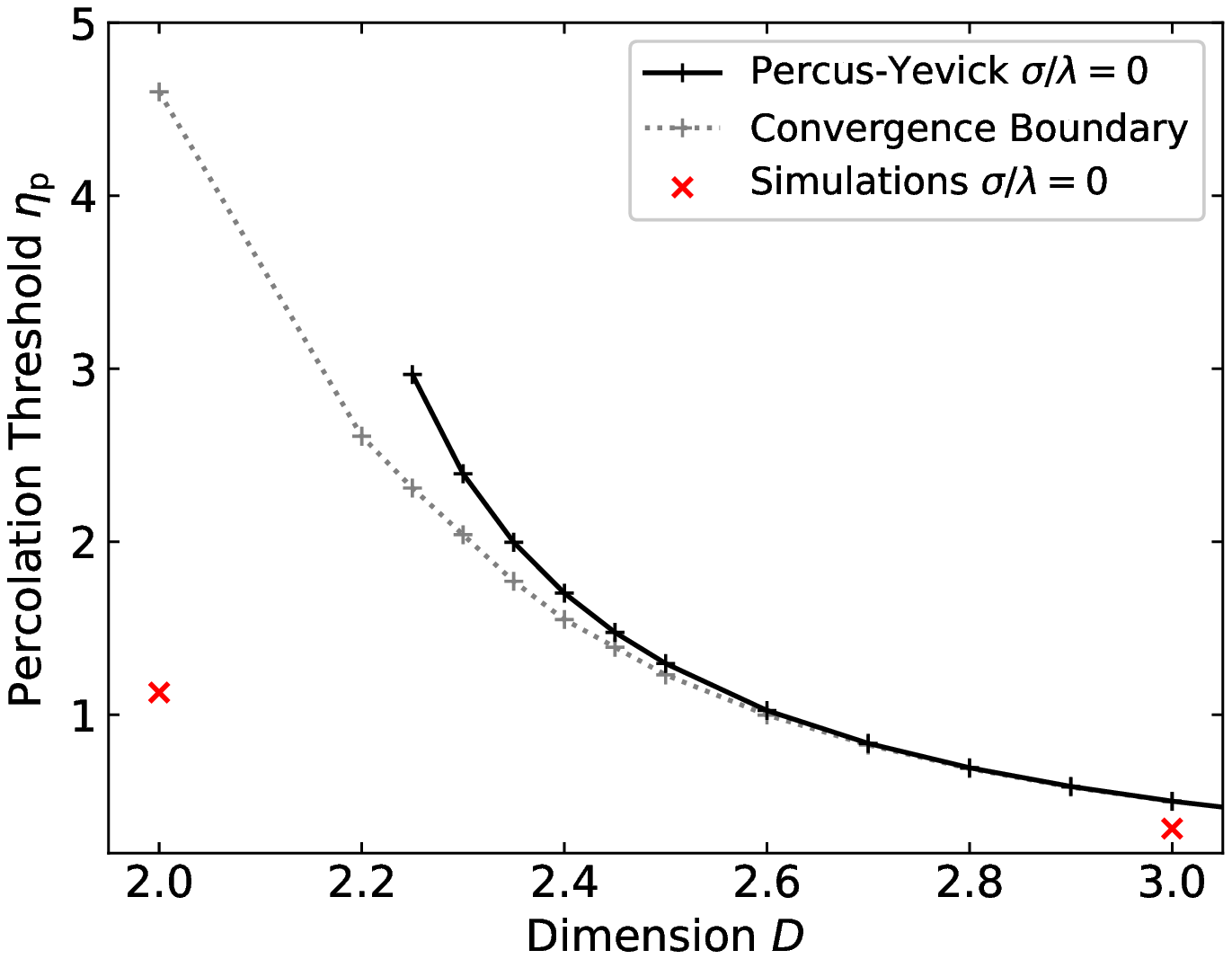}
	\caption{The percolation threshold $\eta_\mathrm{p}$ as function of the spatial dimension $D$ for ideal particles with $\sigma/\lambda = 0$. The percolation threshold is expressed as the number density scaled with the volume of a $D$-dimensional sphere of diameter $\lambda$. Our numerical solutions are shown in black and the convergence boundary of our numerical solver is indicated by the dashed line for $D < 3$.}
	\label{fig:SM:ptconvbound}
\end{figure}

The stability and convergence of our solver decreases significantly upon approaching either a sufficiently high density, or the percolation threshold. Therefore, to be able to obtain an accurate prediction for the percolation threshold and critical exponent, we use a fitting procedure based on the scaling law $S \sim |\eta - \eta_\mathrm{p}|^{-\gamma}$, with $\gamma$ a critical exponent, $\eta$ the scaled density and $\eta_\mathrm{p}$ the percolation threshold. This approach is especially relevant in low dimensional spaces, where the convergence of our solver is poor, even at densities relatively far removed from the percolation threshold. The distance from the convergence boundary of our solver and the percolation threshold is shown in Fig.~\ref{fig:SM:ptconvbound}. Our fitting procedure does not yield a reasonable result for $D < 2.25$, suggesting that we are too far removed from the percolation threshold for the scaling relation to be valid. 

\section*{Validation of procedure}
\begin{figure}[bt]
	\centering
	\includegraphics[width=8.6cm]{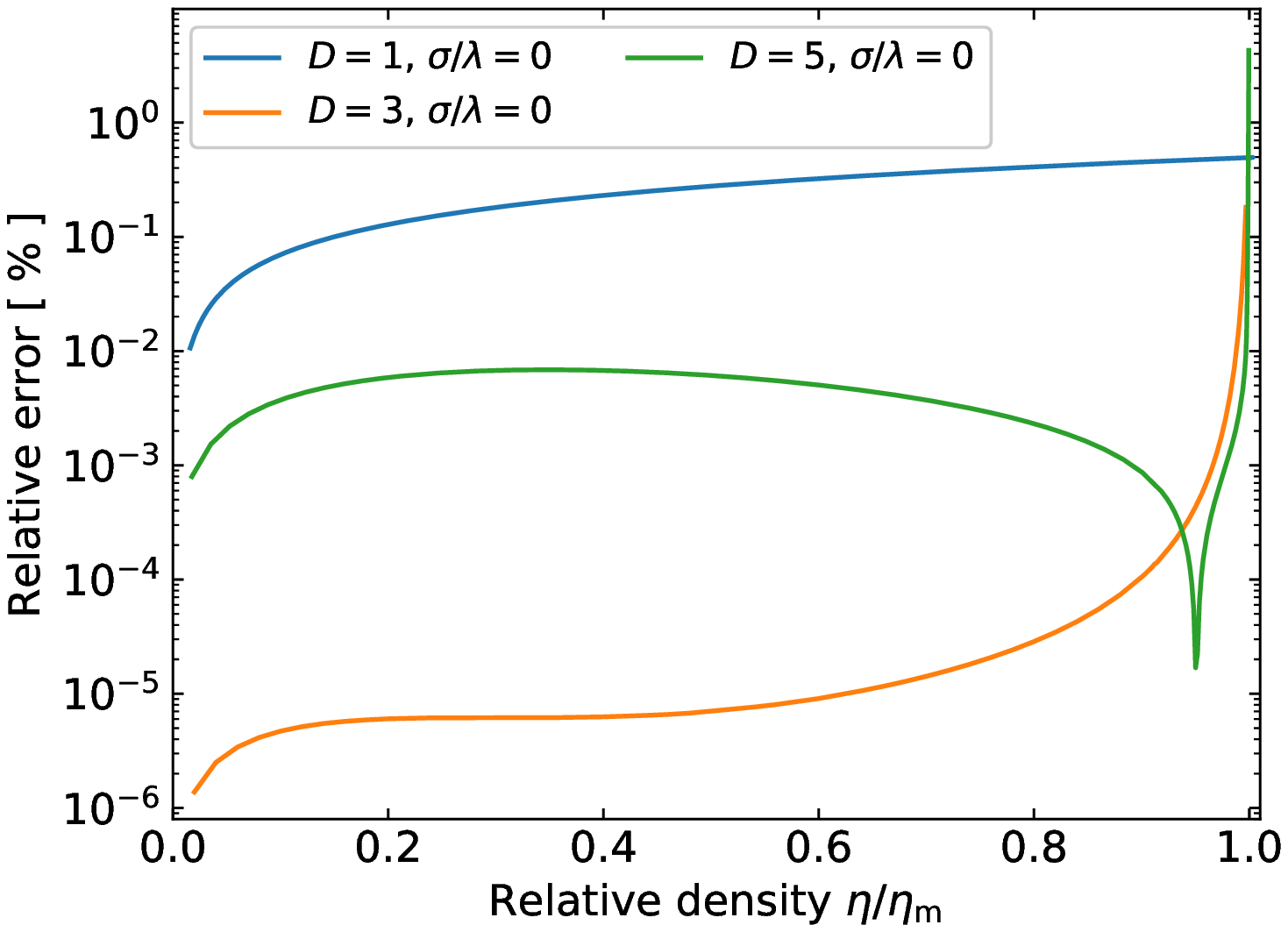}
	\caption{The relative error in the mean cluster size $S$ between our numerical solution and the analytic solution to cPY theory, as function of the relative scaled density $\eta/\eta_\mathrm{m}$, where for $D = 1$ $\eta_\mathrm{m}$ is the highest density our solver converges ($\eta_\mathrm{m} = 25)$, and for $D = 3$ and $D = 5$ the percolation threshold $\eta_\mathrm{p} = 1/2$ and $\eta_\mathrm{p} = 3/2 - 5/6 \sqrt{3}$, respectively.}
	\label{fig:SM:Svsetavalidate}
\end{figure}

To validate that our procedure is correct, we checked that the error of our numerically determined mean cluster size $S = 1 + \rho \widehat{P}(q\to 0)$ remains small with respect to the known analytical results for $D = 1$, $D = 3$ and $D = 5$ for $\sigma/\lambda = 0$. Fig.~\ref{fig:SM:Svsetavalidate} shows this relative error for the mean cluster size. Indeed, we find that away from the percolation threshold the relative error is negligible. Only near the percolation threshold the relative errors increases sharply. This we associate to the pair connectedness function $P(r)$ becoming long-ranged near the percolation threshold. This sudden increase in relative error we associate as the main cause for the large relative errors in $\gamma$.

\section*{Universality class}
\begin{figure}[bt]
	\centering
	\includegraphics[width=8.6cm]{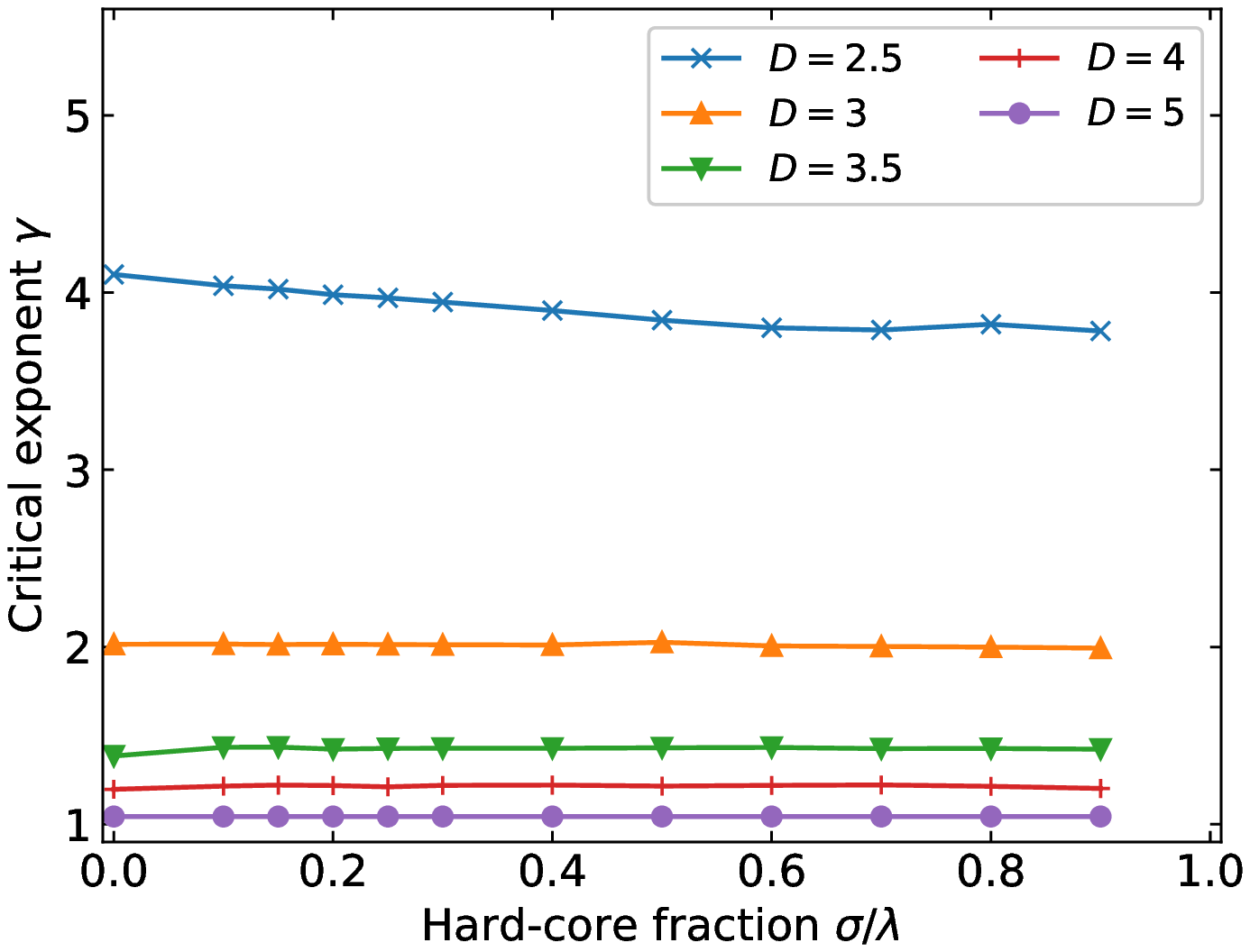}
	\caption{The critical exponent $\gamma$ within cPY theory as function of $\sigma/\lambda$ for $D = 2.5$ (cross), $D = 3$ (triangle-up), $D = 3.5$ (triangle-down), $D = 4$ (plus) and $D = 5$ (dot).}
	\label{fig:SM:gammauniversclassD3}
\end{figure}

The cherry-pit and ideal particle models are generally assumed to be in the same universality class within percolation theory. In Fig.~\ref{fig:SM:gammauniversclassD3}, we highlight that for $D > 2.5$ the critical exponent is (nearly) independent of $\sigma/\lambda$, which strongly suggest that both models indeed fall inside the same universality class, at least within the cPY approach. For  $\sigma/\lambda \to 1$ we find the critical exponent to drop for all dimensions. Hence, this is associated with our numerical approach, and can in be compensated by, \textit{e.g.}, using a finer grid to achieve a higher accuracy.

For $D\leq 2.5$, we find that significant and consistent deviations emerge in $\gamma$. This we directly associate with not having approached the percolation threshold sufficiently close for the presumption that the scaling relation holds to be valid. Since we find the convergence of our solver is significantly worse for $\sigma/\lambda \neq 0$, this shows that these results are less accurate than those for $\sigma/\lambda  = 0$. 

\section*{Analytical solutions}
The mean cluster size in $D = 5$ can be obtained, using the using the known connection within (connectedness) Percus-Yevick theory between the compressibility for hard particles and the mean cluster size for ideal particles \cite{Stell1984a}. Using the analytical expression for the compressibility derived in Ref.~\cite{Leutheusser1984} we can obtain the mean cluster size as
\begin{equation}
S(D = 5) = \frac{9\left(1 + \eta\right)^6}{\left(1-18\eta + 6 \eta^2\right)\left(2-3 \eta + \sqrt{1 - 18 \eta + 6 \eta ^2}\right)^2},
\end{equation}
which diverges at $\eta = \eta_\mathrm{p} = 3/2 - 5/6 \sqrt{3}$. Expanding near the percolation threshold, we find
\begin{equation}
S(\eta \to \eta_\mathrm{p}) = \frac{125\left(7\sqrt{3} -12\right)}{72\left(\eta - \eta_\mathrm{p}\right)},
\end{equation}
hence $\gamma = 1$ in five dimensions.

\section*{Scaled Percolation Threshold}
\begin{figure}[bt]
	\centering
	\includegraphics[width=8.6cm]{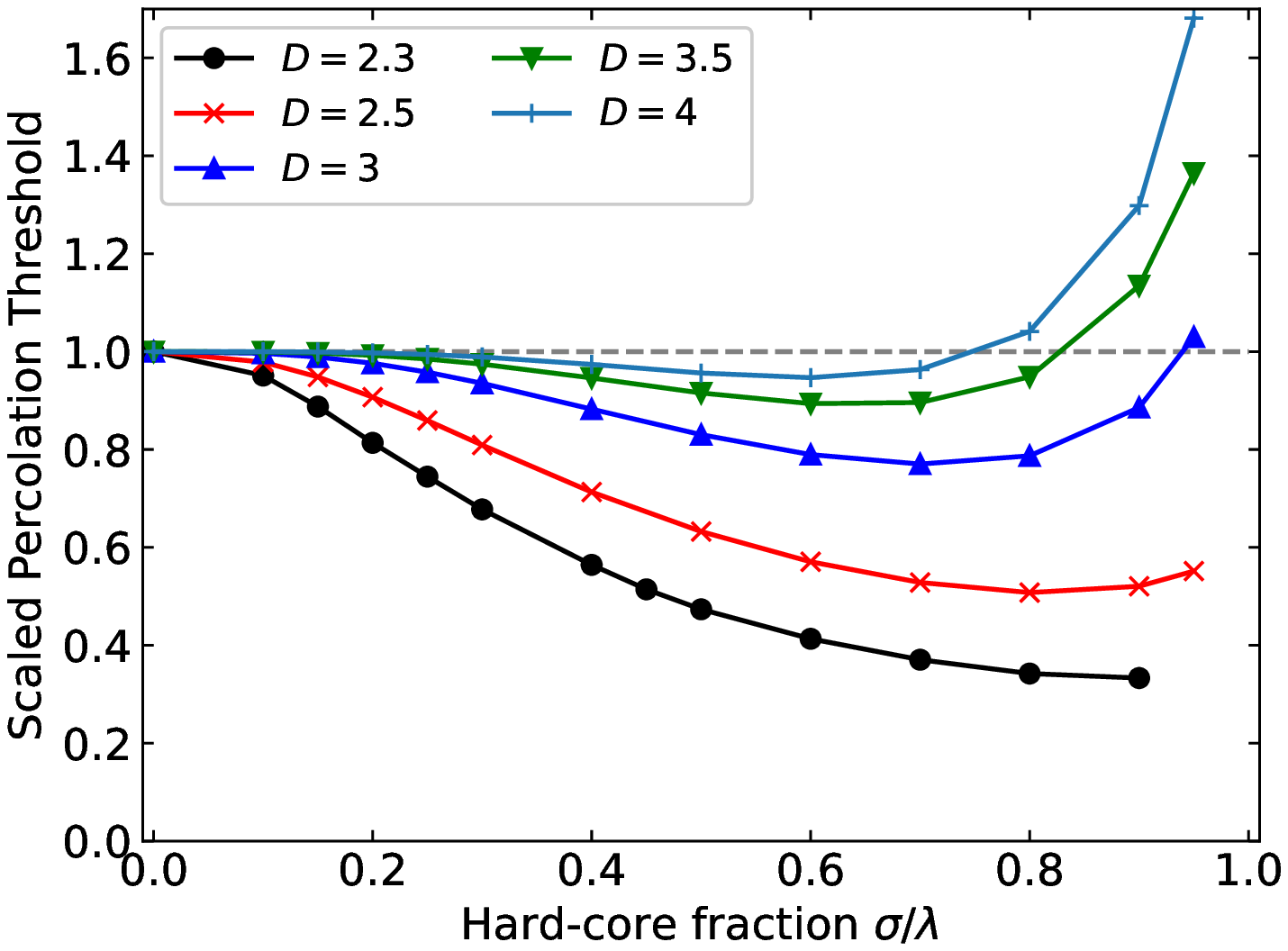}
	\caption{The percolation threshold scaled to the percolation threshold at $\sigma/\lambda = 0$, as function of $\sigma/\lambda$, for $D = 2.3$ (dots), $D = 2.5$ (crosses), $D = 3$ (triangle-up), $D = 3.5$ (triangle-down) and $D = 4$ (plusses).}
	\label{fig:SM:ScaledPT}
\end{figure}

We exemplify the decreasing influence of $\sigma/\lambda$ as function of the dimension in Fig.~\ref{fig:SM:ScaledPT}, where the scaled percolation threshold is plotted as function of $\sigma/\lambda$. With increasing dimension the influence of $\sigma/\lambda$ persists only for relatively large $\sigma/\lambda$.
\bibliography{Bibliography}